\documentclass[aps,prd, twocolumn,floatfix,superscriptaddress,preprintnumbers,
               tightenlines,showpacs,showkeys,nofootinbib,notoccite]{revtex4-2}
\usepackage[utf8]{inputenc}
\usepackage[colorlinks=true,citecolor=blue,linkcolor=blue]{hyperref}
\usepackage[normalem]{ulem}
\usepackage{amsmath,amssymb, mathrsfs}
\usepackage{epsfig}
\usepackage{graphicx}               
\usepackage{url}
\usepackage{color}
\usepackage{slashed}
\usepackage{multirow}
\usepackage{placeins}
\usepackage[dvipsnames]{xcolor}
\usepackage{epstopdf}
\usepackage{soul}
\usepackage{tikz}
\usepackage[capitalise,english]{cleveref}
\usepackage{siunitx}
\usepackage{xspace}
\usepackage{booktabs}
\usetikzlibrary{trees}
\usetikzlibrary{decorations.pathmorphing}
\usetikzlibrary{decorations.markings}

\newcommand\myshade{80}
\colorlet{mylinkcolor}{ForestGreen}
\colorlet{mycitecolor}{Red}
\colorlet{myurlcolor}{violet}

\hypersetup{
  linkcolor  = mylinkcolor!\myshade!black,
  citecolor  = mycitecolor!\myshade!black,
  urlcolor   = myurlcolor!\myshade!black,
  colorlinks = true
}

\definecolor{jblue}{RGB}{20,50,100}
\definecolor{npurple}{RGB} {153, 51, 204}
\definecolor{wred}{RGB}{217,0,56}
\definecolor{white}{RGB}{255,255,255}

\definecolor{korange}{RGB}{235, 80,  43}
\definecolor{korange2}{RGB}{245, 100,  63}
\definecolor{kyelloworange}{RGB}{255, 210,  110}
\definecolor{kyelloworange2}{RGB}{240, 170,  90}
\definecolor{kred}{RGB}{204,  102, 153}
\definecolor{kpurple}{RGB}{153,  61, 190}
\definecolor{kpurplelight}{RGB}{213,  161, 230}

 \definecolor{tobycolour}{rgb}{.5,.0,.5}

\DeclareSIUnit\year{yr}
\DeclareSIUnit\pc{pc}
\DeclareSIUnit\ergs{ergs}
\DeclareSIUnit\msun{\ensuremath{M_\odot}}
\allowdisplaybreaks

\makeatletter
\providecommand*{\diff}%
  {\@ifnextchar^{\DIfF}{\DIfF^{}}}
\def\DIfF^#1{%
  \mathop{\mathrm{\mathstrut d}}%
    \nolimits^{#1}\gobblespace}
\def\gobblespace{%
  \futurelet\diffarg\opspace}
\def\opspace{%
  \let\DiffSpace\!%
  \ifx\diffarg(%
    \let\DiffSpace\relax
  \else
    \ifx\diffarg[%
      \let\DiffSpace\relax
    \else
        \ifx\diffarg\{%
        \let\DiffSpace\relax
      \fi\fi\fi\DiffSpace}

\usepackage{tikz,xcolor,hyperref}

\definecolor{lime}{HTML}{A6CE39}
\DeclareRobustCommand{\orcidicon}{\hspace{-1mm}
	\begin{tikzpicture}
	\draw[lime, fill=lime] (0,0) 
	circle [radius=0.16] 
	node[white] {{\fontfamily{qag}\selectfont \tiny \,ID}};
	\draw[white, fill=white] (-0.0525,0.095) 
	circle [radius=0.007];
	\end{tikzpicture}
	\hspace{-3mm}
}

\foreach \x in {A, ..., Z}{\expandafter\xdef\csname orcid\x\endcsname{\noexpand\href{https://orcid.org/\csname orcidauthor\x\endcsname}
			{\noexpand\orcidicon}}
}

\numberwithin{equation}{section}

\usepackage[compat=1.1.0]{tikz-feynman}
\usepackage{subcaption}

\keywords{}

\begin{document}

\title{\color{black}{\bf Troubles mounting for multipolar dark matter}}

\author{Debajit Bose\orcidA{}}
\email{debajitbose550@gmail.com}
\affiliation{Department of Physics, Indian Institute of Technology Kharagpur, Kharagpur 721302, India}

\author{Debtosh Chowdhury\orcidB{}} 
\email{debtoshc@iitk.ac.in}
\affiliation{Department of Physics, Indian Institute of Technology Kanpur, Kanpur 208016, India}

\author{Poulami Mondal\orcidC{}} 
\email{poulami.mondal1994@gmail.com}
\affiliation{Department of Physics, Indian Institute of Technology Kanpur, Kanpur 208016, India}

\author{Tirtha Sankar Ray\orcidD{}} 
\email{tirthasankar.ray@gmail.com}
\affiliation{Department of Physics, Indian Institute of Technology Kharagpur, Kharagpur 721302, India}

\date{\today}


\begin{abstract}
\noindent In this paper, we revisit the experimental constraints on the multipolar dark matter that has derivative coupling to the visible sector mediated by the Standard Model photon. The momentum dependent interaction enables them to be captured efficiently within massive celestial bodies boosted by their steep gravitational potential. This phenomena makes compact celestial bodies as an efficient target to probe such type of dark matter candidates. We demonstrate that a synergy of the updated direct detection results from DarkSide-50 and LUX-ZEPLIN together with IceCube bounds on high energy solar neutrinos from dark matter capture disfavour the viable parameter space of the dipolar dark matter scenario. Whereas, for the anapole dark matter scenario, a narrow window survives that lies within the reach of prospective heating signals due to the capture of dark matter at cold neutron stars.
\end{abstract}

\maketitle

\section{Introduction}
\label{sec:intro}

Weakly Interacting Massive Particles (WIMPs) that interact to the Standard Model (SM) sector via a derivative coupling, hold significant interest for its rich phenomenological consequences. Unlike generic WIMP models, the momentum dependent couplings soften the correlation between dark matter (DM) annihilation in the early universe and the direct detection cross-section at the present day. In this article, we consider scenarios where the DM has a momentum dependent photon mediated coupling to the visible sector through electric and magnetic dipole or anapole form factors \cite{1958JETP....6.1184Z,Pospelov:2000bq,Sigurdson:2004zp,Masso:2009mu,Ho:2012bg}. This provides a more viable alternative to the millicharged DM where it couples directly to the charged monopole term \cite{Holdom:1985ag,Goldberg:1986nk,Cheung:2007ut}.

To constrain the parameter space of these models we study the impact of their capture at celestial bodies and subsequent observable signatures \cite{Press:1985ug,Gould:1987ju,Gould:1987ir,Garani:2018kkd}. This is of particular importance for scenarios considered in this article where the momentum dependent scattering responsible for their capture is boosted by the gravitational focusing near massive celestial objects. This can be easily assessed by noting that the typical escape velocity on the surface of the neutron star is $\sim 0.6 c$ which can be contrasted with the typical escape velocity at Earth $(\sim 8\times10^{-4} c)$, making the capture mechanism a relevant handle to explore these class of models. We demonstrate that this gravitational boost factor together with the large density of particles inside these compact stars allow us to probe deeper into the parameter spaces of these scenarios. It is known that dark matter scattering through light mediators are predominantly soft resulting in dramatic drop of the capture probabilities \cite{Dasgupta:2020dik}. Interestingly, the effective operators arising from an underlying derivative coupling considered in this work facilitate sufficient momentum transfer in dark matter scattering processes, thereby increasing their capture rate in celestial bodies. Within this paradigm, we explore two signal topologies to constrain the multipolar DM models. The DM captured in compact quantum stars would annihilate and deposit most of its energy content within the stellar environment \cite{Dasgupta:2019juq,Dasgupta:2020dik,Joglekar:2020liw,Leane:2020wob,Bell:2020jou,Bell:2021fye,Anzuini:2021lnv,Alvarez:2023fjj}. This would result in stellar heating and may be observable in colder neutron stars \cite{Baryakhtar:2017dbj,Raj:2017wrv,Bell:2018pkk,Joglekar:2019vzy,Garani:2019fpa,Maity:2021fxw,Bell:2023ysh}. We consider the possibility of observing such neutron star heating at infrared telescopes like James Webb Space Telescope (JWST) with sensitivities of $\mathcal{O}(1000\,$K) to explore these class of DM models. On the other hand, in gaseous main sequence stars, the annihilation products (e.g. $\nu$'s, $\gamma$, etc.) of the captured DM particles may emerge from the stars acting as direct messenger of DM capture \cite{Kopp:2009et,IceCube:2012ugg,Super-Kamiokande:2015xms,IceCube:2016yoy,Garani:2017jcj,Maity:2023rez}. We study the high energetic solar neutrino signatures from DM capture and put constraints on the parameter space of multipolar DM using the existing results from IceCube and DeepCore \cite{IceCube:2016dgk}.

Owing to its coupling to the photons, these class of multipolar DM models are severely constrained by the null results at various direct detection experiments. However, due to the momentum dependence of the couplings and the low ambient velocity of the DM in our neighbourhood, the direct detection bounds are softer. Direct detection limits from the PICO-60, XENON-1T and other detectors have been discussed in \cite{DelNobile:2014eta,Geytenbeek:2016nfg,Kang:2018rad,PICO:2022ohk,Ibarra:2022nzm}. In this article, we update the direct detection limits for anapole and dipole DM models using the most recent published results from DarkSide-50 \cite{DarkSide-50:2022qzh} and LUX-ZEPLIN (LZ) \cite{LZ:2022ufs}. We find that a combination of updated direct detection bounds and high energy solar neutrino constraints from captured DM disfavour the theoretically allowed and phenomenologically viable dipolar DM scenario for DM masses ranging from $1\,$GeV-$10^6\,$GeV. A tuned parameter space survives for the anapole DM framework that remains within the reach of sensitivity of neutron star heating signatures through DM capture.

The article is organised as follows: In section~\ref{sec:anapole_DM}, we have demonstrated the interaction Lagrangian for anapole dark matter and calculate the total annihilation cross-section. We then use the Boltzmann equation to obtain relic density for the anapole dark matter. We have studied the direct detection bounds with specifications of DarkSide-50 and LZ experiments in section~\ref{subsec:DM_DD}. In section~\ref{subsec:DM_capture}, we have discussed DM capture inside a neutron star and the subsequent dark heating.  Next we explore, the framework to analyse the neutrinos from captured DM annihilation inside the Sun. We reviewed our results for dipole dark matter models in section~\ref{sec:dipole_DM}. Finally, we conclude in section~\ref{sec:conclusion}. In appendix~\ref{app:scat_comp}, we compare scattering cross-sections for the different multipolar dark matter scenarios discussed in this work and the kinematics of the scattering processes inside a neutron star are discussed in appendix~\ref{app:scat_kinematics}.
%
%
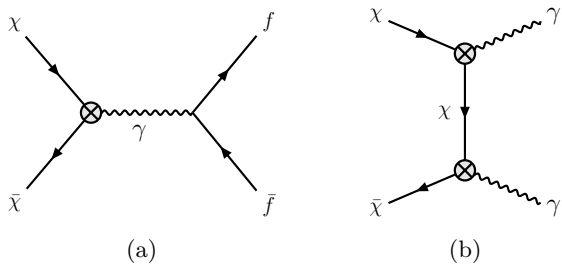
\begin{figure}[t]
    \centering
    \begin{subfigure}{0.485\columnwidth}
        \centering
        \scalebox{0.675}{
        \begin{tikzpicture}[baseline=(current bounding box.center)]
        \begin{feynman}[every crossed dot={/tikz/fill=gray!20,/tikz/inner 				sep=4pt,/tikz/line width=1.25pt}]      
                \vertex [crossed dot] (a4) {};
                \vertex [left = 1.5cm of a4] (a3) ;
                \vertex [above = 1.5cm of a3] (a1) {\large $\chi$};
                \vertex [below = 1.5cm of a3] (a2) {\large $\bar{\chi}$};
                \vertex [right = 2cm of a4] (a5) ;
                \vertex [right = 1.5cm of a5] (a6) ;
                \vertex [above = 1.5cm of a6] (a7) {\large $f$};
                \vertex [below = 1.5cm of a6] (a8) {\large $\bar{f}$};
                \vertex [right = 0.75 cm of a2] (c1);
                \vertex [above = 0 cm of c1] (c2);
                \vertex [above = 2.5 cm of c1] (c3);
                \vertex [right = 3.4 cm of c2] (c4);
                \vertex [right = 3.4 cm of c3] (c5);
                \vertex [right = 1.7cm of c2] (c6);
                \vertex [above = 1.1cm of c6] (c7) {\Large $\gamma$};

                \diagram{
                        (a1) -- [fermion, very thick] (a4) [blob];
                        (a4) -- [fermion, very thick] (a2);
                        (a4) -- [photon, very thick] (a5);
                        (a8) -- [fermion, very thick] (a5);
                        (a5) -- [fermion, very thick] (a7);   
                    };
            \end{feynman}
        \end{tikzpicture}
        }
        \caption{}
        \label{sfig:ann_diagram_a}
    \end{subfigure}
    \begin{subfigure}{0.485\columnwidth}
        \centering
        \scalebox{0.675}{
        \begin{tikzpicture}[baseline=(current bounding box.center)]
        \begin{feynman}[every crossed dot={/tikz/fill=gray!20,/tikz/inner 				sep=4pt,/tikz/line width=1.25pt}]
            	    \vertex (a3) {};
                \vertex [left = 1.75cm of a3] (a1) {\large $\chi$};
                \vertex [right = 1.75cm of a3] (a2) {\Large $\gamma$};
                \vertex [crossed dot, below = 0.75cm of a3] (d1) {};
                \vertex [below = 3.8cm of a3] (b3) {};
                \vertex [crossed dot, above = 0.75cm of b3] (d2) {};
                \vertex [left = 1.75cm of b3] (b1) {\large $\bar{\chi}$};
                \vertex [right = 1.75cm of b3] (b2) {\Large $\gamma$};
                \vertex [below = 1.9cm of a3] (c1);
                \vertex [left = 0.15cm of c1] (c2) {\large $\chi$};
        
                \diagram{
                        (a1) -- [fermion, very thick] (d1);
                        (d1) -- [photon, very thick] (a2);
                        (d1) -- [fermion, very thick] (d2);
                        (d2) -- [fermion, very thick] (b1);
                        (d2) -- [photon, very thick] (b2);  
                    };
        \end{feynman}
        \end{tikzpicture}
        }
        \caption{}
        \label{sfig:ann_diagram_b}
    \end{subfigure}
    \caption{Feynman diagrams of the annihilation processes that are crucial for the relic density calculation.}
    \label{fig:ann_diagram}
\end{figure}
%
%
%
\section{Anapole dark matter}
\label{sec:anapole_DM}

In this section, we will discuss the experimental status of the anapole (AP) DM model \cite{1958JETP....6.1184Z,Pospelov:2000bq,Ho:2012bg}. The dim-6 interaction Lagrangian of a spin-$\frac{1}{2}$ DM particle with photons due to its anapole moment can be expressed as \cite{Ho:2012bg}
\begin{equation}
\label{eq:Lag_AP}
\mathcal{L}_{\rm anapole} = \frac{1}{\Lambda_1^2} \bar{\chi} \gamma_{\mu} \gamma_5 \chi \partial_{\nu} F^{\mu \nu},
\end{equation}
where $\chi$, $F_{\mu \nu}$ are the Majorana DM field, electromagnetic field strength tensor, respectively and $\Lambda_1$ is the EFT cut-off scale. The derivative coupling leads to a momentum dependence in the photon mediated interaction of the DM with the visible sector. While our focus remains on models where DM interacts solely with SM photons, in a more comprehensive framework, it may also interact with hypercharge gauge bosons which can give rise to intriguing phenomena, such as $Z$-boson resonances \cite{Arina:2020mxo,Choi:2024uva}.
%
%
\subsection{Relic Density}
\label{subsec:relic}

We will consider the usual thermal origin of the DM i.e. the particles are in thermal equilibrium with the primordial soup and after freeze-out the relic of the DM particles saturates the measured density of DM. The relic abundance of the DM can be obtained by solving the relevant Boltzmann equation \cite{Kolb:1990vq}
\begin{equation}
\label{eq:BE}
\frac{d Y_{\chi}}{dz} = - \frac{z s \langle \sigma_{\rm ann}v \rangle}{H \left( m_{\chi} \right)} \left( Y_{\chi}^2 - Y_{\rm eq}^2 \right),
\end{equation}
where $Y_{\chi} = n_{\chi}/s$, $n_{\chi}$, $s$ being the DM number density and total entropy density of the universe, respectively. In Eq. \eqref{eq:BE}, $z = m_{\chi}/T$, $m_{\chi}$ is the DM mass and $T$ is the photon temperature, $H(m_{\chi})$ is the Hubble expansion rate and $\langle \sigma_{\rm ann} v \rangle$ is the thermally averaged annihilation cross-section defined in \cite{Gondolo:1990dk}. In the case of anapole DM, the DM annihilation to the fermions are shown in Fig. \ref{sfig:ann_diagram_a}, which is the relevant channel for relic calculations. One can easily read off from Eq. \eqref{eq:Lag_AP}, that the process $\chi \bar{\chi} \rightarrow \gamma \gamma$ is kinematically allowed but forbidden at tree level, because $\partial_\nu F^{\mu \nu} = \partial^\mu (\partial_\nu A^\nu) -\partial^2 A^\mu= 0 $ for on-shell photons. Thus the amplitude of Fig.~\ref{sfig:ann_diagram_b} identically goes to zero for anapole DM scenario. The differential cross-section of the annihilation rate is given by
\begin{equation} \label{eq:sigma_ann_AP}
\begin{split}
\left( \frac{d \sigma_{ \chi \bar{\chi} \rightarrow f \bar{f} }}{d t} \right)_{\rm AP} & = \frac{1}{16 \pi s \left(s - 4 m_{\chi}^2 \right)} \times \frac{8 \pi \alpha_e}{\Lambda_1^4} \left[ 2m_f^4 + 2m_{\chi}^4 \right. \\
 & \hspace*{0.5cm} + s^2 + 2st + 2t^2 - 4m_f^2 \left( m_{\chi}^2 + t \right) \\ 
 & \hspace*{0.5cm} \left. - 4m_{\chi}^2 \left( s+t \right) \right],
\end{split}
\end{equation}
where $\alpha_e$ is the fine-structure constant, $m_f$ is the mass of the fermion, $s$ and $t$ are the Mandelstam variables. We have solved Eq. $\eqref{eq:BE}$ numerically by extracting the thermally averaged annihilation cross-section utilising Eq. $\eqref{eq:sigma_ann_AP}$ and obtained the relic contour by matching it with the observed value obtained in \emph{Planck} 2018 \cite{Planck:2018vyg}.
%
%
%
\begin{figure*}[t]
\begin{center}
\includegraphics[width=0.55\textwidth]{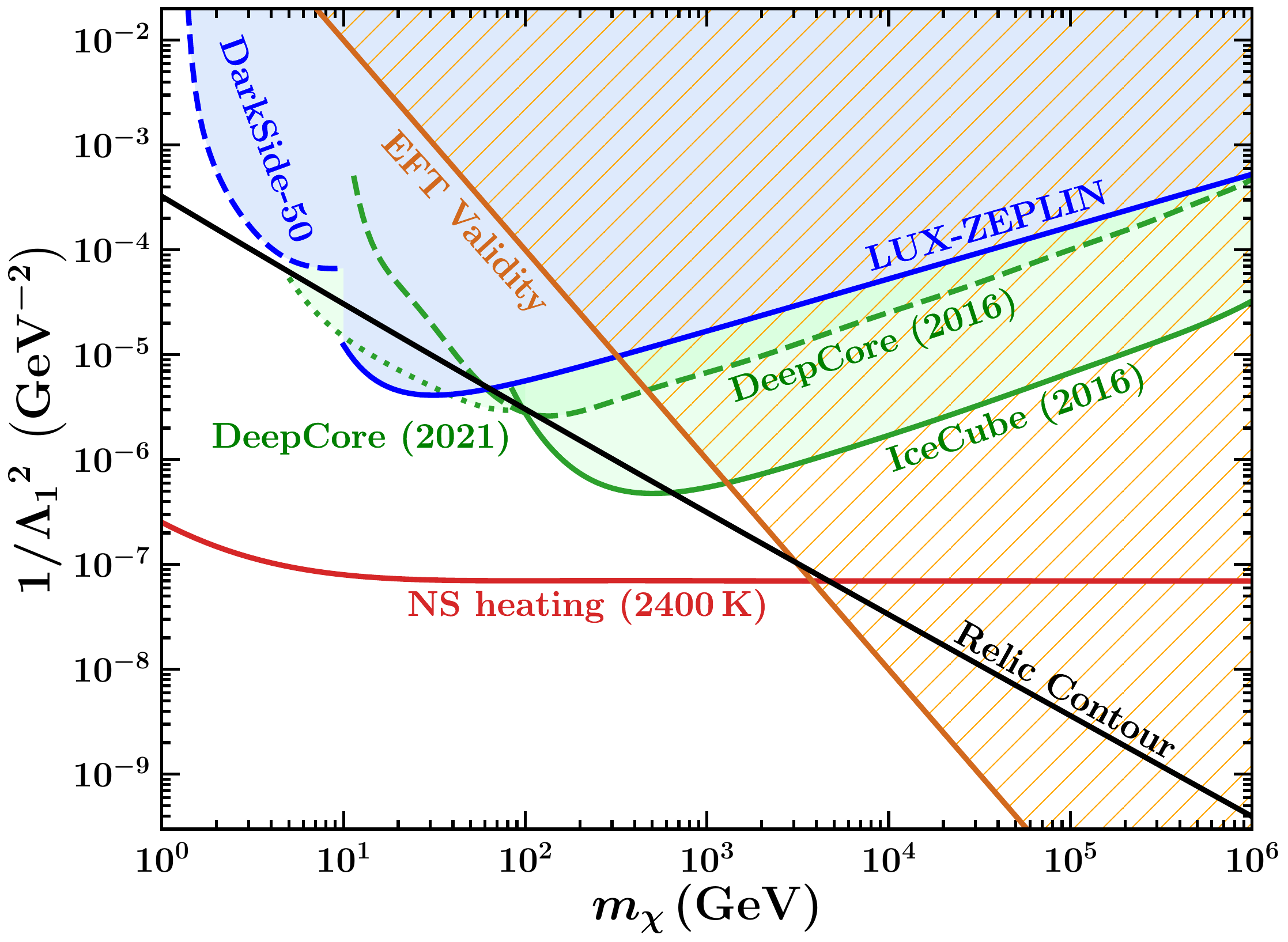}
\caption{The sensitivities of DM coupling for different searches for anapole DM (AP). The black line shows the contour for which the DM relic abundance satisfies \emph{Planck} 2018 \cite{Planck:2018vyg}. Direct detection limits are shown in dashed blue line for DarkSide-50 \cite{DarkSide-50:2022qzh} and using solid blue for LZ \cite{LZ:2022ufs} experiment. We have shown the bounds obtained from analysing the high energy solar neutrinos from captured DM annihilation and compared those with the observations of DeepCore and IceCube \cite{IceCube:2016dgk} by green dashed and solid lines,  respectively. The green dotted curve is derived through a comparative analysis with the upper limits of the annihilation rate of captured DM, employing the latest DeepCore data \cite{IceCube:2021xzo}. Projected limits from dark heating of old cold NS has been shown in red solid line where the NS surface temperature is considered to be $2400 \, $K. In the orange hatched region EFT validity of the considered anapole DM model is no longer maintained.}
\label{fig:AP_limits} 
\end{center}
\end{figure*} 
%
%
%
\subsection{Direct Detection}
\label{subsec:DM_DD}

At direct detection experiments, the ambient DM particles scatter off the nuclei at state-of-the-art experimental facilities and the recoil rates are measured at the detectors. The differential recoil rate of DM scatters off a target nucleus is given by \cite{Schumann:2019eaa}
\begin{equation}
\label{eq:dR_dEnr}
\frac{dR}{dE_{\rm nr}} = \frac{\eta_{\rm exp}}{m_k} \left( \frac{\rho_0}{m_{\chi}} \right) \int_{u_{\rm min}}^{u_{\rm esc}} d u_\chi \, u_\chi \, f(u_\chi) \, \frac{d \sigma}{d E_{\rm nr}},
\end{equation}
where $m_k$ is the mass of the target nucleus, $\eta_{\rm exp}$ is the exposure time of the detector in kg-days and $\rho_0$, $m_{\chi}$ are the local DM density and DM mass, respectively. In Eq. \eqref{eq:dR_dEnr}, $f(u_\chi)$ is the normalized velocity distribution profile of DM and $\frac{d \sigma}{d E_{\rm nr}}$ is the differential scattering cross-section. The minimum velocity of the DM required to produce a recoil event is given by \cite{Schumann:2019eaa}
\begin{equation}
\label{eq:u_min}
u_{\rm min} = \sqrt{ \frac{E_{\rm nr} \, m_T}{2 \beta^2_{\chi,k} } },
\end{equation}
%
with $\beta_{\chi, k}$ is the reduced mass of DM and the nucleus $k$. In Eq. \eqref{eq:dR_dEnr}, the Milky Way escape velocity is given by $u_{\rm esc} = 528 \, \rm km/s$ \cite{deason2019local}. The nuclear recoil rates for anapole model has been obtained with \texttt{WIMpy\_NREFT} \cite{2021ascl.soft12010K}. We update the constraints on the parameter space of anapole DM with recent results from DarkSide-50 \cite{DarkSide-50:2022qzh} and LUX-ZEPLIN (LZ) \cite{LZ:2022ufs}. A brief discussion about these experiments are now in order.
%
%
\subsubsection{DarkSide-50 Experiment}
\label{subsubsec:DS50}

The DarkSide-50 experiment is located in the Hall C of the Gran Sasso National Laboratory in Italy. DarkSide-50 collaboration use liquid Ar as the detector material with an exposure of 12 ton-days \cite{DarkSide-50:2022qzh}. The collaboration has reported the differential recoil events in terms of number of ionised electrons $(N_{\rm e})$ with null results. These electron numbers can be related to the nuclear recoil energy though the relation $N_{\rm e} = Q_y(E_{\rm nr}) \, E_{\rm nr},$ where $Q_y(E_{\rm nr})$ is the liquid Ar ionisation response in terms of nuclear recoil, which has been extracted from \cite{DarkSide-50:2022qzh}. The differential recoil rate in terms of $N_{\rm e}$ is given by
\begin{equation}
\label{eq:dR_dNe}
\frac{dR}{dN_{\rm e}} = \left( \frac{1}{ Q_y(E_{\rm nr}) + E_{\rm nr} \, Q_y'(E_{\rm nr}) } \right) \, \frac{dR}{dE_{\rm nr}},
\end{equation}
where $Q_y'(E_{\rm nr})$ is the differentiation of the ionisation response function with respect to the nuclear recoil energy ($E_{\rm nr}$). For the study of exclusion bounds, we have used the following $\chi^2$-analysis
\begin{equation}
\label{eq:chi^2_DS}
\chi^2_{\rm DS} = \sum_i \left( D_i - B_i - N_{i, \rm pred} \right)^2 / \sigma^2_i,
\end{equation}
where $B_i$, $D_i$, $N_{i, \rm pred}$ are the background, signal and DM induced events in the $i$-th bin and $\sigma_i$ is the variance of the same bin. These values are taken from Ref.~\cite{DarkSide-50:2022qzh}. In the DM mass vs. coupling plane, we put bounds at $90 \%$ confidence level.
%
%
\subsubsection{LUX-ZEPLIN Experiment}
\label{subsubsec:LZ}

The LZ experiment is situated at a depth of 4850 ft within the Davis Cavern, located at the Sanford Underground Research Facility in Lead, South Dakota, USA. The LZ collaboration uses 5.5 ton of Xe as target material and the exposure time for the search is 60 live days and have reported a null result \cite{LZ:2022ufs}. The recoil events are given in terms of electron recoil energy $E_{\rm ee}$. The nuclear recoil and the equivalent electron recoil energy is related by $E_{\rm ee} = Y(E_{\rm nr}) \, E_{\rm nr},$ where $Y(E_{\rm nr})$ is the quenching factor \cite{Essig:2018tss}. The differential recoil rate in terms of equivalent electron recoil energy $(E_{\rm ee})$ is given by \cite{Essig:2018tss}
\begin{equation}
\label{dR_dEee}
\frac{dR}{dE_{\rm ee}} = \left( \frac{1}{ Y(E_{\rm nr}) + E_{\rm nr} Y'(E_{\rm nr}) } \right) \, \frac{dR}{dE_{\rm nr}},
\end{equation}
where $Y'(E_{\rm nr})$ is the differentiation of the quenching factor with respect to $E_{\rm nr}$. To obtain the limits, we have utilized the $\chi^2$-analysis for the Poisson distributed data \cite{Fairbairn:2008gz,Maity:2022exk}
\begin{equation}
\label{eq:chi^2_LZ}
\chi^2_{\rm LZ} = 2 \sum_i \left[ N_{i, \rm pred} + B_i - D_i + D_i \, {\rm log} \left( \frac{D_i}{N_{i, \rm pred} + B_i} \right) \right],
\end{equation}
where $B_i$, $D_i$ and $N_{i, \rm pred}$ are the background, signal and DM predicted events in the $i$-th energy bin. The background and signal events have been extracted from \cite{LZ:2022ufs}. The value of $\chi^2_{\rm LZ}$ defined above depends on the DM mass and coupling with the SM particles. We scan the coupling for fixed DM mass and put bounds at $90 \%$ confidence level as previously explained.
%
%
\subsection{Dark Matter Capture}
\label{subsec:DM_capture}

In this section, we will discuss possible constraints from the capture of anapole DM inside the celestial objects. As a celestial body wanders through a galactic halo of DM particles, it can attract the DM in their gravitational well. The DM particles can scatter with the SM constituents of the celestial body and can loose sufficient energy to become captured within the celestial environment. These accumulated DM particles can annihilate which can increase the stellar luminosity or can produce observable signals at Earth based experiments. As the compact stars have more density than any detector at Earth, indirect searches related to capture can probe much deeper region in the DM parameter space. However, in the case of the DM model discussed in this article, there is an additional advantage due to the momentum dependent interaction topology. As the DM particles are accelerated due to the gravitational focusing at the stars, the DM velocity approaches the escape velocity of the star which boosts the scattering cross-section of the DM and SM particles as discussed in appendix~\ref{app:scat_comp}. The models of DM with derivative coupling to the visible sector considered here has additional significance in terms of the potency of DM capture in constraining its parameter space. We will now explore the two possible complimentary signals from DM capture that will be utilised to constrain the parameter space of the anapole DM.
%
%
\subsubsection{Heating Signature inside NS}
\label{subsubsec:NS_heating}

Owing to its quantum nature, the compact neutron stars are extremely dense with very high escape velocity $\sim 0.6c$, making them ideal for capture signals of DM scenarios like the anapole DM. After thermalization inside the NS, the DM particles annihilate to SM states where the annihilated products are expected to get trapped within the dense neutron star environment given the SM particles have mean free path smaller than the radius of the star. The captured DM can heat the neutron stars both through scattering with the and annihilation processes \cite{Baryakhtar:2017dbj,Raj:2017wrv,Bell:2018pkk,Joglekar:2019vzy,Garani:2019fpa,Maity:2021fxw,Bell:2023ysh}. For DM mass in the range $1-10^6\,$GeV, it can be assumed to be captured within the neutron star following a single scattering event\footnote{ As discussed in \cite{Dasgupta:2020dik}, the DM scattering with nucleons becomes increasingly soft as the mediator becomes light. However, the derivative coupling in the multipolar DM models can alter the kinematics leading to an enhanced energy transfer as discussed in appendix~\ref{app:scat_kinematics}, and thus validating the assumption of capture in compact stars following a single scattering event. }\cite{Raj:2017wrv,Bell:2020jou}. Within this framework, the combined kinetic and annihilation heating for a neutron star with mass, $M_{\rm NS} = 1.5 \, M_{\odot}$ and radius, $R_{\rm NS} = 10 \, \rm km$ can be estimated as \cite{Bell:2023ysh}
\begin{equation}
\label{eq:T_eff}
T_{\rm eff} = 2410 \, f_{\rm cap}^{ \frac{1}{4} } \, \left( \frac{\rho_{\chi}}{ 0.4 \, {\rm GeV/cm^3} } \right)^{ \frac{1}{4} } 
\end{equation}
where $\rho_{\chi}$ is the DM density at the vicinity of the NS which is fixed to the local neighbourhood value of 0.4 $\rm GeV/cm^3$ and the velocity of the NS has been considered to be same as the Sun.
In Eq. \eqref{eq:T_eff}, $f_{\rm cap}$ is the fraction of DM particles that are captured within the NS and is given by
\begin{equation}
\label{eq:f_cap}
f_{\rm cap} \sim {\rm Min} \left[ \sum_i \dfrac{\sigma_{\chi i}}{\sigma_{ {\rm th},i }}, 1  \right],
\end{equation}
where $\sigma_{\chi i}$ is the DM scattering cross-section with target $i$ and $\sigma_{ {\rm th},i }$ is the NS threshold cross-section for target $i$ that can be written as, $\sigma_{ {\rm th},i } = \pi \, R_{\rm NS}^2 / N_i$, where 
$m_i$, $N_i$ are the mass of the target nucleon and number of scattering target inside a neutron star. To calculate the capture fraction defined in Eq. \eqref{eq:f_cap}, we have considered contribution from both protons and neutrons. The photon has interactions through monopole, dipole, and also higher order multipoles with a SM fermion. However, as the neutron is electrically neutral, the leading order contribution to the photon coupling comes from the dipole term that is produced at the loop level. But for the protons, there exists a tree level coupling between the proton and the photon. In this analysis, we have not considered the sub-leading proton interactions via dipole coupling with the photons. The sub-leading contribution is suppressed by the nucleon mass and also by the low proton fraction inside the neutron star. Within this approximation, the contributions from the neutron and the proton are of the same order for the anapole DM model considered in this section. We take a conservative approach and taken the proton fraction inside a NS to be $1\%$ with the understanding that any increase in the proton fraction will make our limits more stringent \cite{Bell:2019pyc}. In the relativistic limit, the differential scattering cross-section of DM with proton and neutron are respectively given by
\begin{gather}
\label{eq:dsigma_chi_p_AP}
\begin{split}
\left( \frac{d \sigma_{ \chi p \rightarrow \chi p }}{ d \cos \theta } \right)_{\rm AP} & = \frac{1}{ 32 \pi s } \times \frac{8 \pi \alpha_e}{\Lambda_1^4} \left[ 2 \left( m_p^4 + m_{\chi}^4 \right) \right. \\
 & \hspace*{0.5cm} + 2 s^2 + 2 s t + t^2 - 4 m_p^2 \left( m_{\chi}^2 + s \right) \\
 & \hspace*{0.5cm} \left. - 4 m_{\chi}^2 \left( s + t \right) \right],
\end{split}
\\
\label{eq:dsigma_chi_n_AP}
\begin{split}
\left( \frac{d \sigma_{ \chi n \rightarrow \chi n }}{ d \cos \theta } \right)_{\rm AP} & = \frac{1}{ 32 \pi s } \times \frac{4 \mu_n^2 t}{\Lambda_1^4} \left[ m_n^2 \left( 2s + 2t - 6 m_{\chi}^2 \right) \right. \\
 & \hspace*{0.5cm} \left. - m_n^4 - \left( m_{\chi}^2 - s \right)^2 - st \right].
\end{split}
\end{gather}
where $m_p$, $m_n$ are the masses of the proton and neutron, respectively. If the neutron star is old enough, it can cool down to $\mathcal{O}(1000 \, \rm K)$ which can be explored by infrared telescopes like JWST \cite{Gardner:2006ky}, Thirty Meter Telescope (TMT) \cite{Crampton:2008gx},  and  European Extremely Large Telescope (E-ELT) \cite{Maiolino:2013bsa}. As discussed in ref. \cite{Chatterjee:2022dhp}, NS temperature $\gtrsim 2400\,$K can be detected in our local vicinity with JWST. So, we have estimated sensitivities in the DM coupling by considering the dark kinetic and annihilation heating of NS to be $2400\,$K.
%
%
\subsubsection{Neutrinos from Sun}
\label{subsubsec:solar_ann}

The complimentary possibility is that the annihilation products from the captured DM is capable of escaping from the gaseous stars and produce observable neutrino signals at the Earth based observatories. Owing to its proximity, the Sun provides the ideal site for analysing such signals. However, if we consider the DM to dominantly annihilate into long-lived mediators, we can also detect charged particles and photons from the annihilation of captured DM inside the Sun or any other celestial body or in a distribution of stars can be detected at Earth based observatories like IceCube, Fermi Large Area Telescope (LAT) and the Alpha Magnetic Spectrometer (AMS), etc. \cite{Batell:2009zp,Feng:2016ijc,Leane:2017vag,Leane:2021ihh,Leane:2021tjj,Bose:2021yhz,Bose:2021cou,Bhattacharjee:2022lts,Nguyen:2022zwb,Chen:2023fgr,Acevedo:2023xnu,Bhattacharjee:2023qfi}. However, as for the DM models we have considered in the article does not have any dark mediator, we will restrict ourselves to the neutrino observations which can escape from the solar interior. Being a main sequence star, the capture rate within the Sun needs to be calculated considering its chemical composition and it is given by \cite{Busoni:2017mhe,Garani:2017jcj}
\begin{equation}
\label{eq:C_Sun}
\begin{split}
C_{\odot} & = \sum_k \left( \frac{\rho_0}{m_{\chi}} \right) \int_0^{R_\odot} 4 \pi r^2 \, dr \\
 & \hspace*{0.5cm} \times \int_0^{u_{\rm esc}} du_{\chi} \frac{f_{v_{\odot}}(u_{\chi})}{u_{\chi}} \, w(r) \, \Omega_k^-(w),
\end{split}
\end{equation}
where the sum is over all possible nuclei present in the solar interior. $w(r)$ is the velocity of the DM particles at a distance $r$ from the center of the sun which can be read as $w(r) = \sqrt{u_{\chi}^2 + v_{\rm esc}^2(r)}$, $v_{\rm esc}(r)$ is the escape velocity of the Sun at the same location. $f_{v_{\odot}}(u_{\chi})$ is the velocity distribution profile of DM particles in the rest frame of the Sun and is given by \cite{Garani:2017jcj}
\begin{equation}\label{eq:f(u)_sun}
\begin{split}
f_{v_{\odot}}(u_{\chi}) & = \sqrt{\frac{3}{2 \pi}} \left( \frac{u_{\chi}}{v_{\odot} \, v_d} \right) \left[ \exp \left\{ - \frac{3 \left( u_{\chi} - v_{\odot} \right)^2}{2 v_d^2 } \right\} \right. \\
 & \hspace*{0.5cm} \left. - \exp \left\{ - \frac{3 \left( u_{\chi} + v_{\odot} \right)^2}{2 v_d^2} \right\} \right],
\end{split}
\end{equation}
where $v_{\odot}$ is the velocity of the Sun in the rest frame of the DM halo and $v_d$ is the dispersion velocity. In this work, we have considered the Maxwell-Boltzmann profile for velocity distribution\footnote{For its deviation and their consequences see \cite{Bose:2022ola}.}. In Eq. \eqref{eq:C_Sun}, the term $\Omega_k^-(w)$ measures the capture probability of DM particles coming with a velocity $w(r)$ that interacts with a nucleus $k$, which can be expressed as \cite{Geytenbeek:2016nfg}
\begin{equation}
\label{eq:Omega_k}
\Omega_k^-(w) = n_k(r) \, w(r) \int_{\frac{m_\chi u_\chi^2}{2}}^{\frac{m_\chi w(r)^2 \zeta_k}{2 \zeta^2_{k,+}}} \frac{d \sigma_k}{d E_{\rm nr}} \, dE_{\rm nr},
\end{equation}
where $n_k(r)$ is the density of nucleus $k$ at radius $r$, $\zeta_k = \left( m_{\chi}/m_k \right)$ and $\zeta_{k,\pm} = \left( \zeta_k \pm 1 \right)/2$. In Eq. \eqref{eq:Omega_k}, $\frac{d \sigma_k}{dE_{\rm nr}}$ is the differential recoil rate of the DM with the nucleus which is given by \cite{DelNobile:2014eta,Geytenbeek:2016nfg,Kang:2018oej}
\begin{equation}\label{eq:recoil_rate_AP}
\begin{split}
\left( \frac{d \sigma_k}{d E_{\rm nr}} \right)_{\rm AP} & = \frac{8 \, \alpha_e \, m_k}{\Lambda_1^4 \, u_{\chi}^2} \left[ Z_k^2 \left( u_{\chi}^2 - \frac{m_k \, E_{\rm nr}}{2 \beta^2_{\chi, k} } \right) \left| F_{\rm E} (E_{\rm nr}) \right|^2 \right. \\
 & \hspace*{0.5cm} \left. + \frac{I_k + 1}{3 I_k} \left(\frac{\mu_k}{\mu_N}\right)^2 \frac{m_k \, E_{\rm nr}}{m_p^2} \, \left| F_{\rm M} (E_{\rm nr}) \right|^2 \right],
\end{split}
\end{equation}
where $Z_k$, $I_k$, and $\mu_k$ are the atomic number, spin, and magnetic moment of the target nucleus, respectively. $\mu_N$ is the nuclear magneton and $F_{\rm E}(E_{\rm nr})$, $F_{\rm M}(E_{\rm nr})$ are the charge and magnetic dipole form factors as a function of $E_{\rm nr}$ which are extracted from \cite{Ibarra:2022nzm}. To calculate the differential recoil events, the magnetic moments and spins of different nuclei have been taken from \cite{STONE200575}. The number of DM particles inside the Sun is dictated by the interplay of DM capture, annihilation and evaporation which is given by
\begin{equation}
\label{eq:DM_density}
\frac{d N_{\chi}}{dt} = C_{\odot} - C_{\rm ann} \, N_{\chi}^2 - C_{\rm evap} \, N_{\chi},
\end{equation}
where $C_{\rm ann}$, $C_{\rm evap}$ are the DM annihilation and evaporation rate inside the Sun. For the DM mass range we are considering i.e. $m_{\chi} > 5 \, {\rm GeV}$, the evaporation rate is numerically insignificant and under equilibrium condition \cite{Garani:2021feo}, we can relate the capture and annihilation rate by
\begin{equation}
\label{eq:Gam_ann}
\Gamma_{\rm ann} = \frac{1}{2} C_{\rm ann} \, N_{\chi}^2 = \frac{C_{\odot}}{2},
\end{equation}
If the DM particles annihilate to SM particles other than photons, the primary or secondary neutrinos from the annihilation can escape from the solar atmosphere and can be detected at Earth based neutrino detectors. The differential neutrino flux reaching at the surface of the Earth can be expressed as
\begin{equation}
\label{eq:diff_neutrino}
E_{\nu}^2 \, \frac{d \phi_{\nu}}{d E_{\nu}} = \frac{\Gamma_{\rm ann}}{4 \pi D_{\odot}^2} \, \times \left( E_{\nu}^2 \, \frac{d N_{\nu}}{dE_{\nu}} \right),
\end{equation}
where $D_{\odot}$ is the distance between the Earth and the Sun, $\frac{d N_{\nu}}{d E_{\nu}}$ is the neutrino spectra per DM annihilation. As in our models, the DM particles annihilate to different final state channels with different branching ratios and all of the channels can contribute to the neutrino spectra, we have taken care of all the channels that can lead to neutrino spectra with proper weightage. The branching fraction of each of the annihilation channels has been calculated using Eq. \eqref{eq:sigma_ann_AP} and the production spectra of the neutrinos are obtained utilizing $\chi \rm{aro} \nu$ \cite{Liu:2020ckq}. These produced neutrinos need to be propagated through the solar interior, vacuum, Earth and its atmosphere to reach the detector. The propagation of the neutrinos have been calculated using \texttt{nuSQuIDS} \cite{Arguelles:2021twb}, where neutrino oscillations, scattering and tau regeneration have been considered and the obtained neutrino spectra has been utilized in Eq. \eqref{eq:diff_neutrino} to calculate the differential neutrino flux at the detector.

In this work, we have utilised the IceCube and DeepCore events published by the IceCube collaboration \cite{IceCube:2016dgk}, where the analysis uses the 3 years data between May, 2011 and May, 2014. The analysed events are muon track-like events and have been studied with respect to cosine of the solar opening angle $(\Psi_{\odot})$ in Fig. 6 of \cite{IceCube:2016dgk}. The DM induced differential event rate in terms of $(\Psi_{\odot})$ is given by \cite{Maity:2023rez}
\begin{equation}
\label{eq:N_Psi}
\begin{split}
N_{\Psi_{\odot}} & = 2 \, T \int_{E_{\nu}^{\rm min}}^{E_{\nu}^{\rm max}} A_{\rm eff}(E_{\nu}) \frac{d \phi_{\nu}}{d E_{\nu}} \, \frac{1}{\sqrt{2 \pi} \sigma_{\psi}} \\ 
 & \hspace*{0.5cm} \times \exp \left[ - \frac{\left( \cos(\Psi_{\odot}) - 1 \right)^2}{2 \sigma_{\psi}^2} \right] dE_{\nu},
\end{split}
\end{equation}
where $A_{\rm eff}(E_{\nu})$ is the effective area of the detector that has been extracted from Fig. 4 of \cite{IceCube:2016dgk} and $T = 532 \, {\rm days}$ is the exposure time for which the Sun is below the horizon. In Eq. \eqref{eq:N_Psi}, the flux is calculated for the muon flavour neutrino and anti-neutrino only, as the analysis done in Ref.~\cite{IceCube:2016dgk} considered muon track-like events. The dispersion $(\sigma_{\psi})$ in Eq. \eqref{eq:N_Psi} is given by
\begin{equation}
\label{eq:sigma}
\sigma_{\psi} = \left| \frac{\sqrt{2} \left( 1 - {\rm cos}\left[ \Delta \Psi(E_{\nu}) \right] \right)}{2 \, {\rm erf^{-1}}\left(0.5\right)} \right|,
\end{equation}
where $\Delta \Psi(E_{\nu})$ be the angular resolution of the detector shown in Fig. 4 of \cite{IceCube:2016dgk}. The exclusion limits have been obtained through $\chi^2$-analysis where $\chi^2$ is defined as
\begin{equation}
\label{eq:chi^2_sun}
\chi^2 = \sum_i \left( D_i - B_i - N_{i, \Psi_{\odot}} \right)^2 / \sigma^2_i 
\end{equation} 
where $B_i$, $D_i$, $N_{i, \Psi_{\odot}}$ and $\sigma_i$ are the background, observed data, DM induced events and variance in the $i$-th bin of $\Psi_{\odot}$. These data have been extracted from Ref.~\cite{IceCube:2016dgk}. We have obtained upper limits for each of the models by iterating the DM coupling for a fixed DM mass at $95\%$ confidence level. Recently, the IceCube collaboration has analysed 6.75 years of DeepCore data and provided upper bounds on the annihilation rate of solar captured DM for masses $5\,$GeV-$100\,$GeV for different SM final states \cite{IceCube:2021xzo}. We compare the annihilation rate for different SM channels with Eq. \eqref{eq:Gam_ann} multiplied with proper branching ratios. We find that the $\tau^+\tau^-$ channel imposes most stringent limit, as illustrated in Fig. \ref{fig:AP_limits}.
%
%
%
\begin{figure*}[!]
\centering
\begin{subfigure}{0.45\textwidth}
\centering
\includegraphics[width=1\linewidth]{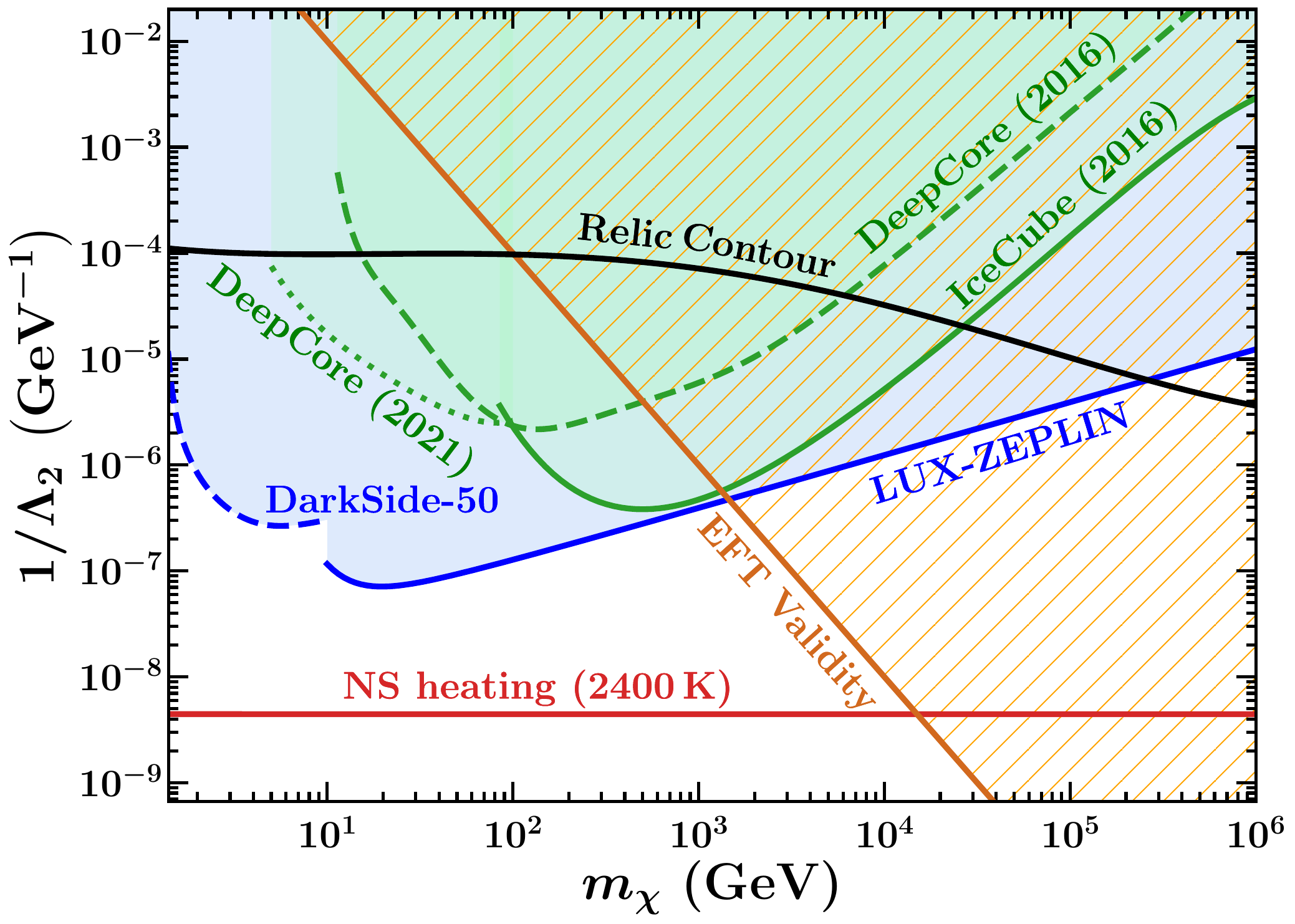} 
\caption{}
\label{sfig:MDM_limits}
\end{subfigure}
\begin{subfigure}{0.45\textwidth}
\centering
\includegraphics[width=1\linewidth]{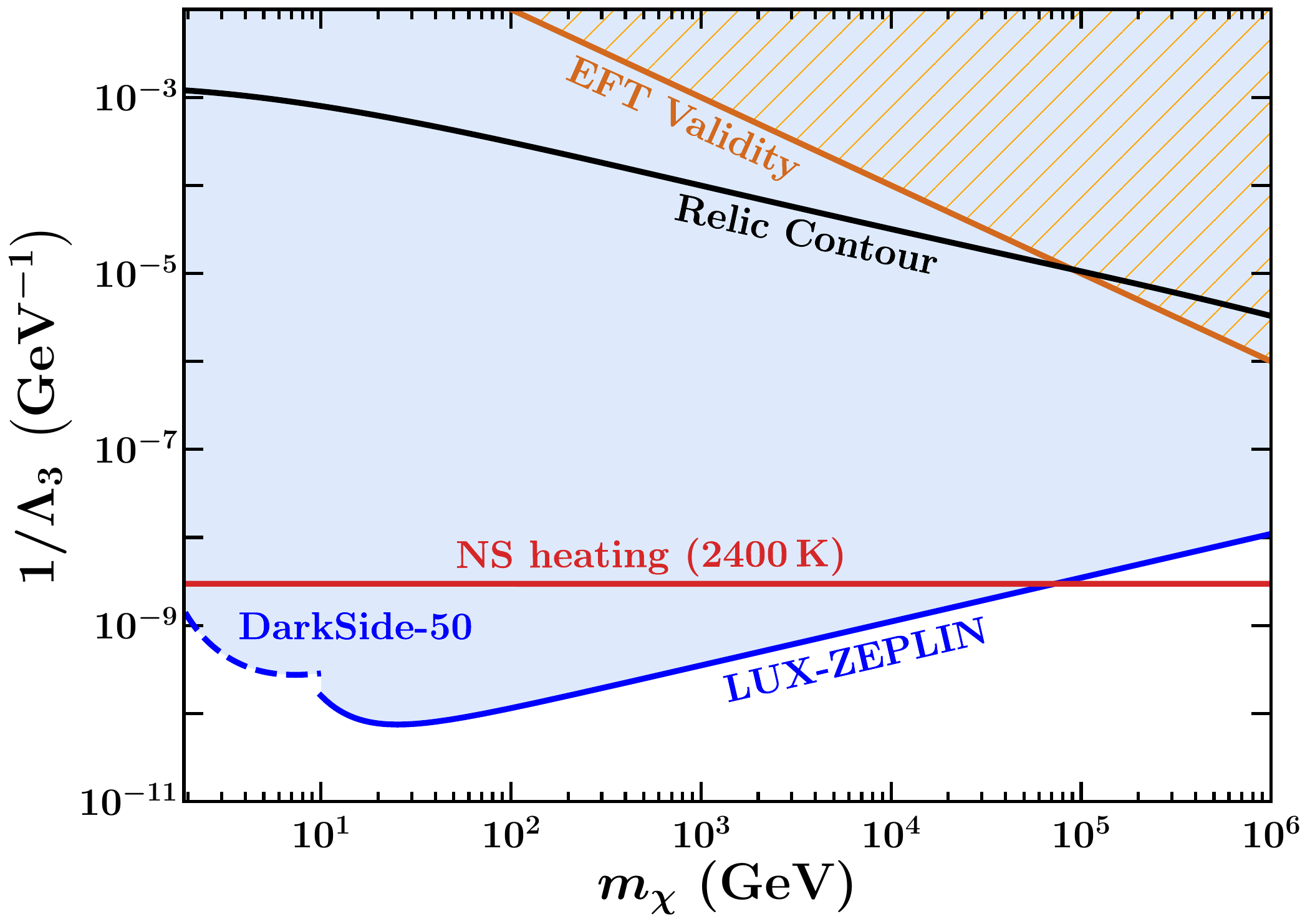} 
\caption{}
\label{sfig:EDM_limits}
\end{subfigure}
\caption{Same as Fig. \ref{fig:AP_limits}, but for magnetic dipole (left) and electric dipole (right) DM models.}
\label{fig:dipole_limits}
\end{figure*}
%
%
%
\subsection{Results}
\label{subsec:results}

In Fig. \ref{fig:AP_limits}, we present the exclusion limits from the synergy of direct and indirect detection experiment on the parameter space of the anapole DM. The black line in Fig. \ref{fig:AP_limits} denotes the relic contour of the concerned model by solving Boltzmann equation described in section \ref{subsec:relic} and matching the abundance with the \emph{Planck} 2018 observation \cite{Planck:2018vyg}. The brown line in the Fig. \ref{fig:AP_limits} depicts the contour satisfying $\Lambda_1 = m_{\chi}$ beyond which the calculations using an effective theory cannot be trusted. In Fig. \ref{fig:AP_limits}, the direct detection limits have been shown with blue solid lines. The DarkSide-50 provides more stringent limit for DM mass $\sim 10 \ {\rm GeV}$, beyond which the LZ limits become more stringent. We have analysed the indirect detection bounds from the signatures of annihilating DM considering the combined analysis of Fermi-LAT and Major Atmospheric Gamma Imaging Cherenkov (MAGIC) \cite{MAGIC:2016xys} telescopes and the latest observations from High Energy Stereoscopic System (H.E.S.S.) \cite{HESS:2022ygk}. However, for the considered DM mass range, the indirect detection bounds constraint the parameter space upto $1 / \Lambda_1^2 \sim 10^{-5} \, {\rm GeV^{-2}}$ and remains sub-dominant compared to the direct detection bounds. We do not show them in Fig.~\ref{fig:AP_limits} to reduce clutter. The bounds obtained by analysing the neutrino signatures from solar captured DM are shown in green colour, where the dashed and solid lines are obtained by utilizing the DeepCore and IceCube data, respectively \cite{IceCube:2016dgk}. We have also shown the limits with green dotted line obtained by comparing with the upper limits outlined in latest DeepCore analysis \cite{IceCube:2021xzo}. In Fig. \ref{fig:AP_limits}, we have also shown the projected limits obtained from the dark heating of a old cold NS with red solid line where the surface temperature of the NS is considered to be $2400 \,$K which can be detected at JWST in the near future \cite{Chatterjee:2022dhp}.

We find that the updated direct detection results and the exclusion contours from the high energetic solar neutrinos from captured DM leave a narrow window of allowed parameter space that is both theoretically consistent and phenomenologically viable. We further demonstrate that this region lies within the sensitivity range for the neutron star heating experiments to explore in the near future.
%
%
\section{Dipole dark matter}
\label{sec:dipole_DM}

Next we turn our attention to the class of models where the DM couples to the electric dipole (ED) and magnetic dipole moment (MD) of the visible sector through a momentum dependent coupling with the photon. The Lagrangian of the models are given by \cite{Sigurdson:2004zp, Masso:2009mu}
\begin{gather}
\label{eq:Lag_dipole}
\mathcal{L}_{\rm dipole} = \underbrace{ \frac{1}{\Lambda_2} \bar{\chi} \sigma_{\mu \nu} \chi F^{\mu \nu} }_\text{magnetic} + \underbrace{ \frac{i}{\Lambda_3} \bar{\chi} \sigma_{\mu \nu} \gamma_5 \chi F^{\mu \nu} }_\text{electric},
\end{gather}
where $\Lambda_2$, $\Lambda_3$ are EFT cut-off scale of the corresponding models.

We repeat the analysis done in the previous section to constrain the parameter space of these models. To keep the analysis generic, we have treated the electric and magnetic dipole model separately. Any specific combination can be easily constrained from the results presented here. Now we systematically enumerate the main analytic expressions central to our analysis. The differential annihilation cross-sections of the DM to fermions for both the models are given by
\begin{gather}
\label{eq:sigma_ff_MD}
\begin{split}
\left( \frac{d \sigma_{ \chi \bar{\chi} \rightarrow f \bar{f} }}{d t} \right)_{\rm MD} & = \frac{1}{16 \pi s \left(s - 4 m_{\chi}^2 \right)} \times \frac{64 \pi \alpha_e}{\Lambda_2^2 s} \left[ 2m_{\chi}^2 t \right. \\ 
 & \hspace*{0.5cm} + 2m_{\chi}^2 s + m_f^2 \left( s + 2t + 2m_{\chi}^2 \right) - m_{\chi}^4 \\
 & \hspace*{0.5cm} \left. - m_f^4 - t \left( s+t \right) \right],
\end{split}
\\
\label{eq:sigma_ff_ED}
\begin{split}
\left( \frac{d \sigma_{ \chi \bar{\chi} \rightarrow f \bar{f} }}{d t} \right)_{\rm ED} & = \frac{1}{16 \pi s \left(s - 4 m_{\chi}^2 \right)} \times \frac{64 \pi \alpha_e}{\Lambda_3^2 s} \left[ 2m_{\chi}^2t \right. \\ 
 & \hspace*{0.5cm} + m_f^2 \left( s + 2t - 2m_{\chi}^2 \right) - m_{\chi}^4 - m_f^4 \\
 & \hspace*{0.5cm} \left. - t \left( s+t \right) \right],
\end{split}
\end{gather}
The differential cross-sections for the photon annihilations are same for both the electric and magnetic dipole models and is given by
\begin{equation}\label{eq:sigma_gg_MD}
\begin{split}
\left( \frac{d \sigma_{ \chi \bar{\chi} \rightarrow \gamma \gamma }}{d t} \right)_{\rm MD(ED) } & = \frac{1}{16 \pi s \left(s - 4 m_{\chi}^2 \right)} \times \frac{256}{\Lambda_{2(3)}^4 \left( m_{\chi}^2-t \right)} \\
 & \hspace*{0.5cm} \times \frac{1}{\left( m_{\chi}^2 - s -t \right)} \times \left[ 4m_{\chi}^6 \left( s+t \right) \right. \\
 & \hspace*{0.5cm} \left. + 4m_{\chi}^2t \left( s+t \right)^2 - t^2 \left( s+t \right)^2 \right. \\
 & \hspace*{0.5cm} \left. - m_{\chi}^4 \left( s^2 + 10st + 6t^2 \right) - m_{\chi}^8 \right].
\end{split}
\end{equation}
Both of these annihilation channels are utilised in Eq. \eqref{eq:BE} to solve the Boltzmann equation for computing the present day DM relic density. 

The differential recoil rate of DM with a target nuclei $(k)$ which are given by \cite{DelNobile:2014eta, Geytenbeek:2016nfg}
\begin{gather}
\label{eq:recoil_rate_MD}
\begin{split}
\left( \frac{d \sigma}{d E_{\rm nr}} \right)_{\rm MD} & = \frac{4 \, \alpha_e}{\Lambda_2^2 \, u_{\chi}^2} \left[ \frac{I_k + 1}{3 I_k} \left(\frac{\mu_k}{\mu_N}\right)^2 \frac{m_k}{m_p^2} \, \left| F_{\rm M} (E_{\rm nr}) \right|^2 \right. \\
 & \hspace*{0.35cm} + \left. \left\{ \frac{u_{\chi}}{E_{\rm nr}} - \frac{1}{2 m_k} - \frac{1}{m_{\chi}} \right\} \, Z_k^2 \, \left| F_{\rm E} (E_{\rm nr}) \right|^2  \right], 
\end{split}
\\
\label{eq:recoil_rate_ED}
\left( \frac{d \sigma}{d E_{\rm nr}} \right)_{\rm ED} = \frac{4 \, \alpha_e \, Z_k^2}{\Lambda_3^2 \, E_{\rm nr} \, u_{\chi}^2} \, \left| F_{\rm E} (E_{\rm nr}) \right|^2, 
\end{gather}
\noindent
The differential recoil rate is required for direct detection and solar capture rate calculations. For the magnetic dipole moment, the recoil rates for direct detection experiments have been calculated using \texttt{WIMpy\_NREFT} \cite{2021ascl.soft12010K}. However, the same for the electric dipole has been calculated by using Eqs. (\ref{eq:dR_dEnr}$\,$$,$$\,$\ref{eq:recoil_rate_ED}). For the high energy solar neutrino analysis, the capture rates for different nuclei have been calculated using Eqs. (\ref{eq:C_Sun}$\,$$,$$\,$\ref{eq:Omega_k}$\,$$,$$\,$\ref{eq:recoil_rate_MD}) for magnetic dipole case. The solar high energy neutrino constraints on the electric dipole DM scenario are sub-dominant due to overwhelming branching fraction to the photons and has been neglected in this analysis. The differential scattering cross-sections of DM particles with protons inside a NS for dipole models are given by
\begin{gather}
\label{eq:dsigma_domega_chi_p_MD}
\begin{split}
\left( \frac{d \sigma_{ \chi p \rightarrow \chi p }}{ d \cos \theta } \right)_{\rm MD} & = \frac{1}{ 32 \pi s } \times \frac{64 \pi \alpha_e}{\Lambda_2^2 t} \left[ 2m_{\chi}^2 \left( s+t \right) - m_p^4 \right. \\
 & \hspace*{0.5cm} \left. - m_{\chi}^4 + m_p^2 \left( 2m_{\chi}^2 + 2s + t \right) \right. \\
 & \hspace*{0.5cm} \left. - s \left( s+t \right) \right],
\end{split}
\\
\label{eq:dsigma_domega_chi_p_ED}
\begin{split}
\left( \frac{d \sigma_{ \chi p \rightarrow \chi p }}{ d \cos \theta } \right)_{\rm ED} & = \frac{1}{ 32 \pi s } \times \frac{64 \pi \alpha_e}{\Lambda_3^2 t} \left[ t \left( m_p^2-s \right) - \right. \\
 & \hspace*{0.5cm} \left. \left( m_p^2 + m_{\chi}^2 -s \right)^2 \right].
\end{split}
\end{gather}
\noindent
Similarly, the differential scattering cross-sections of DM with the neutrons produced via loop level coupling are given by
\begin{gather}
\label{eq:dsigma_domega_chi_n_MD}
\begin{split}
\left( \frac{d \sigma_{ \chi n \rightarrow \chi n }}{ d \cos \theta } \right)_{\rm MD} & = \frac{1}{ 32 \pi s } \times \frac{4 \mu_n^2}{\Lambda_2^2} \left[ 4 \left( m_n^4 + m_{\chi}^4 \right) - 8 m_{\chi}^2 s \right. \\ 
 & \hspace*{0.5cm} \left. + 8 m_n^2 \left( 3 m_{\chi}^2 - s \right) + \left( 2 s + t \right)^2 \right]
\end{split}
\\
\label{eq:dsigma_domega_chi_n_ED}
\begin{split}
\left( \frac{d \sigma_{ \chi n \rightarrow \chi n }}{ d \cos \theta } \right)_{\rm ED} & = \frac{1}{ 32 \pi s } \times \frac{4 \mu_n^2}{\Lambda_3^2} \left[ 4 m_n^4 - 8 m_n^2 \left( m_{\chi}^2 + s \right) \right. \\ 
 & \hspace*{0.5cm} \left. + \left( 2s + t - 2 m_{\chi}^2 \right)^2 \right].
\end{split}
\end{gather}
Using the above equations, we have obtained the limits from dark heating of a cold NS as described in section \ref{subsubsec:NS_heating}. In Fig \ref{fig:dipole_limits}, we present the limits from direct detection and capture scenario for dipole DM candidates.  In Fig. \ref{sfig:MDM_limits}, we present the status for the magnetic dipole DM and Fig. \ref{sfig:EDM_limits} shows a similar plot for electric dipole DM. The black solid lines are the relic contours obtained by consistent with the relic abundance measured at \emph{Planck} 2018 \cite{Planck:2018vyg}. The direct detection lines are marked by blue lines and the red lines are the projecting limits from NS heating having surface temperature $2400 \,$K. The green dashed and solid lines are the bounds obtained from analysing the solar neutrinos using DeepCore and IceCube data, respectively \cite{IceCube:2016dgk}. The green dotted line is obtained by comparing the annihilation rate of the captured DM with the upper limits derived from the latest DeepCore observations \cite{IceCube:2021xzo}. The direct detection limits shown in Fig. \ref{fig:dipole_limits} are stronger for electric dipole model compared to the magnetic dipole model. This can be traced to the fact that in non-relativistic approximation, the magnetic dipole solely contributes to the spin-dependent interactions with the nucleons, whereas the electric dipole DM receives contributions from both spin-independent and spin-dependent interactions \cite{PhysRevC.89.065501}. As we can also see from Fig. \ref{fig:dipole_limits}, for both types of dipole DM models, the phenomenologically relevant parameter space is already disfavoured by the direct detection experiments. However, there might be an interesting region in the sub-GeV DM mass, where the direct detection lines are much more relaxed and projected limits from dark heating of compact stars might put better constraints.
%
%
%
\begin{figure}[t!]
\begin{center}
\includegraphics[width=0.95\columnwidth]{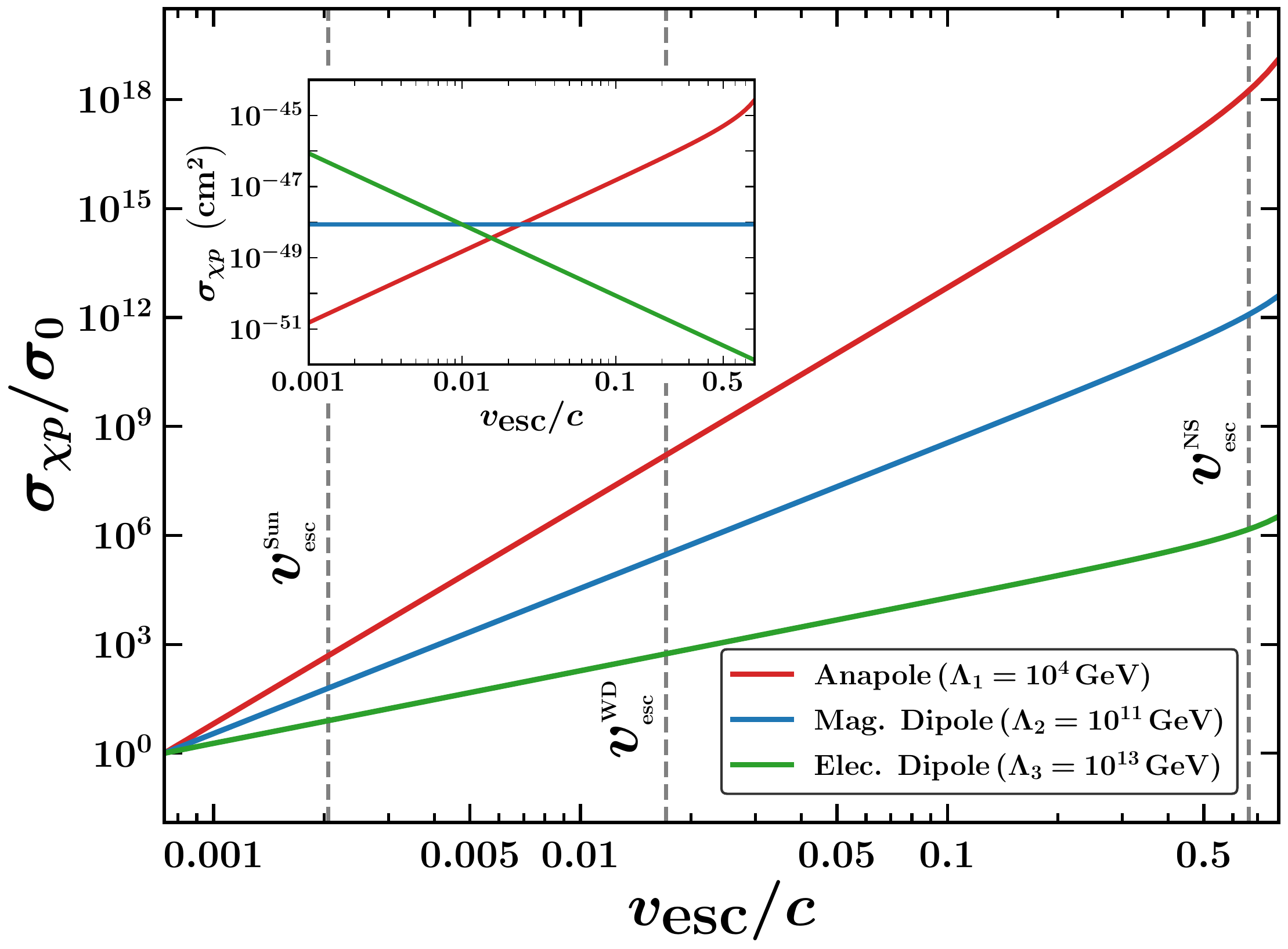}
\caption{Variation of the normalised scattering cross-section as a function of the escape velocity of the celestial object is shown with red, blue, and green lines for anapole, magnetic, and electric dipole DM models, respectively. The grey dashed vertical lines are drawn at $v_{\rm esc}^{\rm Sun}$, $v_{\rm esc}^{\rm WD}$, and $v_{\rm esc}^{\rm NS}$, which are the escape velocities of Sun, white dwarf, and neutron star, respectively.}
\label{fig:scat_comp}
\end{center}
\end{figure}
%
%
%
\section{Conclusions}
\label{sec:conclusion}

In this paper, we make a systematic study of WIMP-like multipolar DM that has a derivative coupling to the photon. The momentum dependence, introduced through the derivative coupling, enhances the scattering rate making them amenable to capture at celestial bodies where DM particles are gravitationally boosted. In this context, we have considered DM capture inside a compact neutron star and obtained projected limits considering the potential detection of such compact stars having surface temperature around $2400\,$K. We further explore the capture of multipolar DM inside the Sun and analyse the constraints arising due to the DM annihilation inside Sun using the observations made at IceCube observatory. We demonstrate that the updated direct detection bounds from the DarkSide-50 and the LZ experiment together with the constraints from IceCube on high energy solar neutrinos, have ruled out the viable parameter space of the electric and magnetic dipole DM models. We further show that in the context of anapole DM, the restricted region of parameter space that survives these bounds lies within the reach of the observatories like the JWST.
%
%
\section*{Acknowledgments}
We thank Tarak Nath Maity, Rohan Pramanick, and Arpan Hait for helpful discussions and valuable comments. DB acknowledges MHRD, Government of India for fellowship. This research of DC is supported by an initiation grant IITK/PHY/2019413 at IIT Kanpur and by a DST-SERB grant SERB/CRG/2021/007579. The research of PM is supported by a DST-SERB grant SERB/CRG/2021/007579.
%
%
\appendix
%
%
\section{Comparison between DM Scattering Rate}
\label{app:scat_comp}

In this section, we will discuss the effect of the gravitational boosting on the DM capture for the momentum dependent DM interaction to the visible sector. To this end, we compare the DM scattering rate in the multipolar DM scenarios with a toy scenario where the interactions are momentum independent. Fig.~\ref{fig:scat_comp} depicts the variation of the scattering cross-section, upto a normalisation $(\sigma_0)$, as a function of escape velocity for all three DM scenarios discussed in this article. For the toy model we have considered, the differential scattering cross-section of DM with the proton is given by
\begin{equation}\tag{A.1}
\label{eq:dsigma_domega_0}
\begin{split}
\left( \frac{d \sigma_{ 0 }}{ d \cos \theta } \right) & = \frac{1}{ 32 \pi s } \times \frac{8 \pi \alpha_e g_{\chi}^2}{t^2} \left[ 2 \left( m_p^2 + m_{\chi}^2 - s \right)^2 \right. \\ 
 & \hspace*{0.5cm} \left. + 2st + t^2 \right],
\end{split}
\end{equation}
where $g_{\chi}$ is the toy model coupling of the DM to the SM sector. From Fig.~\ref{fig:scat_comp}, we see that the scaled scattering cross-section monotonically increases with the escape velocity of the compact celestial body. In the inset of Fig.~\ref{fig:scat_comp}, we show the actual dependency of the scattering cross-sections as a function of $v_{\rm esc}$. Although, it seems that for the electric dipolar DM scenario, the scattering cross-section decreases with increasing $v_{\rm esc}$, it is an artefact of the exchange of a massless mediator. From the plot, we see that for the anapole model the enhancement to the scattering cross-section is maximal compared to the dipole type of interactions. This large boost factor makes the anapole DM the best case scenario in terms of the detection via the capture at a compact celestial body. 

The above discussion have been done considering the DM-proton interaction. Identical boosts are expected if one considers the DM-neutron scatterings. Fig.~\ref{fig:scat_comp} clearly indicates why the capture of multipolar DM is more efficient in compact quantum stars like the neutron star compared to the main sequence stars, owing to their different surface escape velocity. This is the reason behind the fact that potential bounds from neutron stars in Figs.~\ref{fig:AP_limits},\ref{fig:dipole_limits} are more constraining than the corresponding solar limits.
%
%
\section{Kinematics of DM Capture via Light Mediators}
\label{app:scat_kinematics}

\begin{figure}[t!]
\begin{center}
\includegraphics[width=0.95\columnwidth]{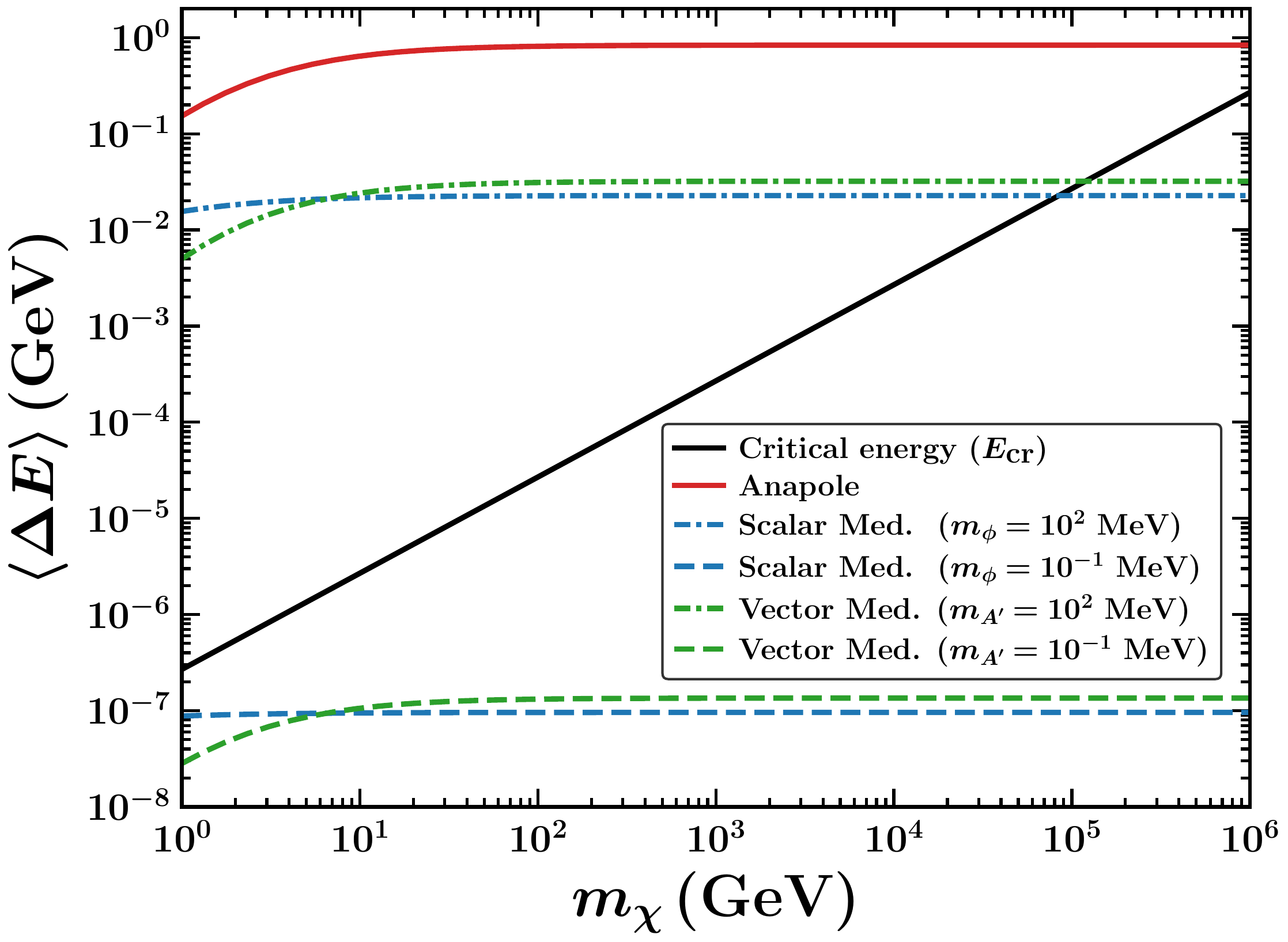}
\caption{Variation of the average energy transfer in a single scattering event within a neutron star with the DM mass for different DM models. The black line represents the critical energy loss required for the DM particles for successful capture. The red line denotes the average energy loss for the anapole DM model, while the blue and green lines correspond to energy losses in scalar and vector-mediated models for different mediator masses as depicted in the legend.}
\label{fig:scat_kinematics}
\end{center}
\end{figure}

In this appendix, we discuss the average energy transfer of the DM particles in a single scattering event within the neutron star. The average energy transfer in a typical scattering of DM particle with a nucleon inside the neutron star is given by \cite{Bell:2018pkk,Bell:2019pyc}
\begin{equation}\tag{B.1}
\label{eq:ER_avg}
\begin{split}
\langle \Delta E \rangle & = \dfrac{\left( 1-B \right) m_{\chi} \zeta_p}{B + 2 \sqrt{B} \zeta_p + B \zeta_p^2} \\ 
 & \hspace*{0.5cm} \times \dfrac{\int_{-1}^{1} d \cos \theta \, \left( 1 - \cos \theta \right) \, \frac{d \sigma}{d \, \cos \theta}}{\int_{-1}^{1} d \cos \theta \, \frac{d \sigma}{d \cos \theta} },
\end{split}
\end{equation}
where $\zeta_p = \left( m_{\chi}/m_p \right)$, $B = \left(1 - 2 G M_{\rm NS}/R_{\rm NS} \right) \approx 0.56$ for typical neutron star under consideration.  This energy transfer from the incident DM to the stellar constituent must cross some critical value $E_{\rm cr}(m_{\chi})$ which is equal to the ambient kinetic energy of the DM, to ensure its capture \cite{Raj:2017wrv}. In Fig. \ref{fig:scat_kinematics}, we have plotted the critical energy loss with black line assuming the average halo velocity of DM to be $\sim 220\,$km/s.
%
%
%
%
%
%

The red line in Fig. \ref{fig:scat_kinematics} depicts the average energy loss for anapole DM model obtained from Eq. \eqref{eq:ER_avg} with the differential scattering cross-section given in Eq. \eqref{eq:dsigma_chi_p_AP}. Similarly, we obtain the blue lines for the scenario where DM scattering is mediated by a scalar mediator leading to a Yukawa-like potential \cite{Dasgupta:2020dik}. The  differential scattering cross-section for such scenario is given by
\begin{equation}\tag{B.2}
\left. \dfrac{d \sigma}{d \cos \theta} \right|_{\phi} = \frac{1}{ 32 \pi s } \times \frac{g_{\chi \phi}^2 g_{\phi f}^2}{\left( t-m_{\phi} \right)^2} \left( 4 m_{\chi}^2 - t \right) \left( 4 m_p^2 - t \right),
\end{equation}
where $g_{\chi \phi}$, $g_{\phi f}$ are respectively the scalar mediator couplings to DM and SM and $m_{\phi}$ is the mass of the scalar mediator. As can be seen from Fig. \ref{fig:scat_kinematics}, for the interactions originated by a scalar mediator, for most of the DM mass range, the energy transfer falls below the threshold as the mediator mass falls below $0.1\,$MeV. Similarly for a vector mediator the differential cross-section is given by
\begin{equation}\tag{B.3}
\begin{split}
\left. \dfrac{d \sigma}{d \cos \theta} \right|_{A'} & = \frac{1}{ 32 \pi s } \times \frac{2 g_{A' \phi}^2 g_{A' f}^2}{\left( t-m_{A'} \right)^2} \left[ 2 \left( m_p^2 + m_{\chi}^2 - s \right)^2 \right. \\ 
 & \hspace*{0.5cm} \left. + 2st + t^2 \right],
\end{split}
\end{equation}
where $g_{\chi A'}$, $g_{A' f}$ epresent the couplings of the vector mediator to DM and SM, respectively, while $m_{A'}$ denotes the mass of the vector mediator. Expectedly,  the vector mediator scenario exhibits a similar dependency on the mediator mass  as  the scalar mediator case.

As can be concluded from Fig. \ref{fig:scat_kinematics}, the anapole DM scenario (and the multipole models by extensions) owing to  the momentum dependent interactions drastically alters the kinematics of the scattering. For all of the parameter space under consideration, the average energy transfer is greater than the critical momentum transfer required for capture.  So, unlike the usual light mediator models, the DM capture within a neutron star through single scattering remains unsuppressed for anapole and dipole DM models due to the presence of derivative couplings.

\vspace*{5cm}
\bibliographystyle{JHEP}
\bibliography{multipole_DM_refs}
\end{document}